\documentstyle[12pt]{article}
\begin{document}
\baselineskip=14pt

\begin{center}
{\Large {\bf A new Dirac-type equation for tachyonic
 neutrinos}}

\vskip 0.8cm
            Tsao Chang \\ 
  Center for Space Plasma and Aeronomy Research\\
  University of Alabama in Huntsville\\
            Huntsville, AL 35899\\
      Email: changt@cspar.uah.edu\\

\vskip 0.5cm
\end{center}

\baselineskip=14pt

  Based on experimental evidences supporting the hypothesis that 
neutrinos might be tachyonic fermions, a new Dirac-type equation 
is proposed and a spin-$\frac{1}{2}$ tachyonic quantum theory is 
developed.  The new Dirac-type equation provides a solution for 
the puzzle of negative mass-square of neutrinos.  This equation 
can be written in two spinor equations coupled together via 
nonzero mass while respecting the maximum parity violation, and 
it reduces to one Weyl equation in the massless limit. Some 
peculiar features of tachyonic neutrino are discussed in this 
theoretical framework.


\newpage
\noindent
{\bf I. INTRODUCTION}

 A model has recently been presented to fit the cosmic ray 
spectrum at $E \approx 1-4$ PeV [1-3] using the hypothesis
that the electron neutrino is a tachyon.  This model yields 
a value for $m^2(\nu_e) \approx -3 eV^2$, which 
is consistent with the results from recent 
measurements in tritium beta decay experiments [4-6]. Moreover,
the muon neutrino also exhibits a negative mass-square [7]. 
However, up to now, there is no satisfactory tachyonic quantum
theory to describe neutrinos as spin-$\frac{1}{2}$ fermions.  

   The negative value of the neutrino mass-square simply means:
     $$  E^2/c^2 - p^2 = m_{\nu}^2 c^2 < 0	\eqno  (1)  $$      
The right-hand side in Eq.(1) can be rewritten as ($-m_s^2 c^2$), 
then $m_s$ has a positive value.  

 Based on special relativity and known as re-interpretation rule, 
tachyon as a hypothetical particle was proposed by Bilaniuk et al. 
in the Sixties [8-10].  For tachyons, the relation of momentum 
and energy is shown in Eq.(1). The negative value on the right-
hand side of Eq.(1) means that $p^2$ is greater than $(E/c)^2$.
The velocity of a tachyons, $u_s$, is greater than speed of light. 
The momentum and energy in terms of $u_s$ are as follows:
    $$	p =  \frac{m_s u_s}{\sqrt{u_s ^2 /c ^2  - 1}},  \quad 		
    	E = \frac{m_s c^2}{\sqrt{u_s ^2 /c ^2  - 1}}     \eqno  (2) 
$$	
where the subscript $s$ means superluminal particle, i.e. tachyon. 

  Any physical reference system is built by subluminal particles 
(such as atoms, molecules etc.), which requires that a reference 
frame must move slower than light.  On the other hand, once a tachyon 
is created in an interaction, its speed is always greater than 
the speed of light.  Neutrino is the most possible candidate for a 
tachyon because it has left-handed spin in any reference 
frame [11,12].  However, anti-neutrino always has right-handed 
spin.  Considering the measured mass-square is negative
for the muon neutrino,  Chodos, Hauser and Kostelecky [11] suggested 
in 1985 that the muon neutrino might be a tachyon.  They also 
suggested that one could test the tachyonic neutrino in high energy 
region using a strange feature of tachyon: $E_\nu$ could be negative 
in some reference frames [13,14].  This feature has been further 
studied by Ehrlich [1-3].  Therefore, it is required to construct a 
tachyonic quantum theory for neutrinos.       

The first step in this direction is usually to introduce an 
imaginary mass, but these efforts could not reach a point for 
constructing a consistent quantum theory.  Some early investigations
of a Dirac-type equation for tachyonic fermions are listed in 
Ref.[15].  An alternative approach was investigated by Chodos et al.
[11].  They examined the possibility that muon neutrino might be
tachyonic fermion.  A form of the lagrangian density for tachyonic 
neutrinos was proposed.  Although they did not obtain a satisfatory 
quantum theory for tachyonic fermions, they suggested that more 
theoretical work would be needed to determine physically acceptable 
theory.  
\\

\noindent
{\bf 2. A NEW DIRAC-TYPE EQUATION}

   In this paper, we will start with a different approach to derive 
a new Dirac-type equation for tachyonic neutrinos.  In order to 
avoid introducing imaginary mass, Eq. (1) can be rewritten as
    $$ E = (c^2p^2 -  m_s^2c^4 )^{1/2}  \eqno  (3)  $$
where $m_s$ is called proper mass.  For instance, 
$m_s(\nu_e)$= 1.6 eV, if taking $m^2(\nu_e)= -2.5 eV^2$ [16].  
We follow Dirac's approach [17],
Hamiltonian must be first order in momentum operator ${\hat p}$:
  $$  \hat E = c {\vec \alpha} \cdot {\hat p} + \beta_s m_s c^2 	
     \eqno (4)  $$
with  ($\hat E = i\hbar \partial /\partial t , {\hat p} = 
-i \hbar \nabla $).  ${\vec \alpha} = (\alpha_1, \alpha_2, 
\alpha_3$) and $\beta_s$ are 4$\times$ 4 matrix, which are defined as
  $$ {\alpha_i} = \left(\matrix{0&\sigma_i\cr
                         \sigma_i&0\cr}\right),  \quad
   \beta_s = \left(\matrix{0&I\cr
                         -I&0\cr}\right)  \eqno (5)   $$
where $\sigma_i$ is 2$\times$2 Pauli matrix, $I$ is 2$\times$2 unit 
matrix.  Notice that $\beta_s$ is a new matrix, which 
is different from the one in the traditional Dirac equation.  
We will discuss the property of $\beta_s$  in a later section.

    When we take square for both sides in Eq.(4), and consider the 
following relations:
 $$ \alpha_i \alpha_j + \alpha_j \alpha_i = 2 \delta_{ij}    $$
      $$	\alpha_i \beta_s + \beta_s \alpha_i =  0  $$
      $$       \beta_s^2 = -1	\eqno    (6)		$$
then the Klein-Gordon equation is derived, and
the relation in Eq.(1) or Eq. (3) is reproduced.  Since Eq.(3) is 
related to Eq. (2), this means $\beta_s$ is a 
right choice to describe neutrinos as tachyons.

 We now study the spin-$\frac{1}{2}$ property of neutrino (or
anti-neutrino) as a tachyonic fermion.  

  Denote the wave function as
  $$ \Psi = \left(\matrix{\varphi ({\vec x},t)\cr
                         \chi ({\vec x},t)\cr}\right) \quad
with \quad
   \varphi = \left(\matrix{\varphi_1\cr
                         \varphi_2\cr}\right), \quad
    \chi = \left(\matrix{\chi_1\cr
                         \chi_2\cr}\right)    $$
the complete form of the new Dirac-type equation, Eq.(4), becomes
$$  \hat E \Psi = c({\vec \alpha} \cdot {\hat p})\Psi + 
           \beta_s m_s c^2 \Psi 	     \eqno (7)  $$
It can also be rewritten as a pair of two-component equations:
$$ i\hbar \frac{\partial \varphi}{\partial t} = -ic 
    \hbar {\vec \sigma} \cdot  \nabla \chi + m_s c^2 \chi   $$
	$$ i\hbar \frac{\partial \chi}{\partial t} = -ic \hbar 
\vec{\sigma} \cdot \nabla \varphi - m_s c^2 \varphi  
\eqno (8)   $$

From the equation (8), the continuity equation is derived:
$$  \frac{\partial \rho}{\partial t} +
           \nabla \cdot {\vec j} = 0  \eqno (9)   $$
and we have
  $$ \rho = \varphi^{\dag} \chi + \chi^{\dag} \varphi ,  \quad
  {\vec j} = c(\varphi^{\dag} {\vec \sigma} \varphi + \chi^{\dag}  
 {\vec \sigma} \chi)    \eqno (10)  $$
where $\rho$ and $\vec j$ are probability density and current; 
$\varphi^{\dag}$ and $\chi^{\dag} $ are the Hermitian adjoint 
of $\varphi$ and $\chi$ respectively.

 Eq.(10) can be rewritten as
  $$ \rho = \Psi^{\dag}  \gamma_5 \Psi ,  \quad
  {\vec j} = c(\Psi^{\dag}  {\vec \Sigma} \Psi) 
     \eqno (11)  $$
where $\gamma_5$ and ${\vec \Sigma}$ are defined as
$$ {\gamma_5} = \left(\matrix{0&I \cr
                         I&0 \cr}\right),  \quad
   {\vec \Sigma} = \left(\matrix{{\vec \sigma}&0 \cr
                   0&{\vec \sigma} \cr}\right)  \eqno (12)   $$
 
Considering a plane wave along the $z$ axis for a right-handed 
particle, the helicity $H = ({\vec \sigma} \cdot {\vec p})/|{\vec p}|
 = 1 $, then the equation (8) yields the following solution:
  $$	 \chi = \frac{cp - m_s c^2}{E} \varphi   \eqno	(13) $$

Indeed, there are four independent bispinors as the solutions of 
Eq.(8). The explicit form of four bispinors are listed in the Appendix. 

 Let us now discuss the property of matrix $\beta_s$ in Eq.(5).  
Notice that it is not a $4 \times 4$ hermitian matrix. As we know,
a non-hermitian Hamiltoian is not allowed for a subluminal particle,
but it does work for tachyonic neutrinos.  Moreover, the 
relation between the matrix $\beta_s$ and the traditional matrix 
$\beta$ is as follows:
$$  \beta_s = \beta \gamma_5    \quad where \quad
   {\beta} = \left(\matrix{I&0 \cr
     0&-I \cr}\right)  \eqno	(14)   $$ 
\\ 

\noindent
{\bf III.  PARITY VIOLATION FOR NEUTRINOS}
  
 In order to compare the new Dirac-type equation Eq.(7) with
the two component Weyl equation in the massless limit, 
we now consider a linear combination of $\varphi$ and $\chi$ :
  $$   \xi= {1 \over {\sqrt 2}} (\varphi + \chi)  , \quad
      \eta = {1 \over {\sqrt 2}} (\varphi - \chi)  \eqno (15)  $$
where $\xi ({\vec x},t)$ and $\eta ({\vec x},t)$ are two-component 
spinor functions. In terms of $\xi$ and $\eta$, Eq.(10) becomes 
  $$ \rho = \xi^{\dag} \xi - \eta^{\dag} \eta ,  \quad
 {\vec j} = c(\xi^{\dag} {\vec\sigma} \xi + \eta^{\dag}{\vec \sigma}
 \eta) \eqno (16)  $$
In terms of Eq.(15), Eq.(8) can be rewritten in Weyl representation: 
$$ i\hbar \frac{\partial \xi}{\partial t} = -ic \hbar {\vec \sigma}
    \cdot \nabla \xi - m_s c^2 \eta   $$
$$ i\hbar \frac{\partial \eta}{\partial t} = ic \hbar {\vec \sigma}
     \cdot  \nabla \eta + m_s c^2 \xi  \eqno (17)   $$
In the above equations, both $\xi$ and $\eta$ are coupled via 
nonzero $m_s$.

  For comparing Eq.(17) with the well known Weyl equation, 
we take a limit, $m_s = 0$, then the first equation in
Eq.(17) reduces to
	$$   \frac{\partial \xi_{\bar{\nu}}}{\partial t} = -c
 {\vec \sigma} \cdot \nabla \xi_{\bar{\nu}}     \eqno (18)   $$
In addition, the second equation in Eq.(17) vanishes because 
$\varphi = \chi$ when $m_s = 0$.

Eq.(18) is the two-component Weyl equation for describing 
anti-neutrinos ${\bar{\nu}}$, which is related to the 
maximum parity violation discovered in 1956 
by Lee and Yang [18,19].  They pointed out that no experiment had 
shown parity to be good symmetry for weak interaction.  Now we see 
that, in terms of Eq.(17), once if neutrino has some mass, no 
matter how small it is, two equations should be coupled together via 
the mass term while still respecting the maximum parity violation. 

 Indeed, the Weyl equation (18) is only valid for antineutrinos 
since a neutrino always has left-handed spin, which is opposite 
to antineutrino.  For this purpose, we now introduce a 
transformation:
        $$   \vec{\alpha}  \rightarrow - {\vec \alpha}
 	\eqno (19) $$
It is easily seen that the anticommutation relations in Eq.(6) remain 
unchanged under this transformation.  In terms of Eq.(19), Eq.(4) 
becomes
$$  \hat E \Psi_\nu = -c ({\vec \alpha} \cdot {\hat p})\Psi_\nu + 
           \beta_s m_s c^2 \Psi_\nu 	     \eqno (20)  $$
Furthermore, ${\vec \sigma}$ should be replaced by
(${-\vec \sigma}$) from Eq.(5) to Eq.(18) for describing a neutrino.
Therefore, the two-component Weyl equation (18) for massless 
neutrino becomes: 
 $$   \frac{\partial \xi_\nu}{\partial t} = c{\vec \sigma} \cdot
           \nabla \xi_\nu      \eqno (21)   $$
 In fact, the transformation (19) is associated with the CPT theorem.
Some related discussions can be found in Ref.[20-22].  As an 
application, Eq.(19) can also be used to discuss the traditional 
Dirac equation in terms of $\Psi_d $ for electrons, which is
$$  \hat E \Psi_d = c ({\vec \alpha} \cdot {\hat p})\Psi_d + 
           \beta m_o c^2 \Psi_d 	 \eqno (22)  $$
where $m_o$ is the rest mass of electron.

  Using the transformation (19), Eq.(22) becomes
$$  \hat E \Psi_a = -c({\vec \alpha} \cdot {\hat p})\Psi_a + 
           \beta m_o c^2 \Psi_a 	 \eqno (23)  $$
where $\Psi_a$ means that ($-\vec{\alpha}$) is adopted in the 
equation.  We suggest that Eq.(23) can be used to describe a 
positron or other subluminal antiparticles with 
spin-$\frac{1}{2}$.\\

\noindent
{\bf IV. SUMMARY AND DISCUSSIONS}

 In this paper, The hypothesis that neutrinos might be tachyonic 
fermions is further investigated. A spin-${\frac{1}{2}}$ 
tachyonic quantum theory is developed on the basis of the new 
Dirac-type equation. It provides a solution for the puzzle 
of negative mass-square of neutrinos.
 
  The matrix $\beta_s$ in the new Dirac-type equation
 is not a $4 \times 4$ hermitian matrix.  However, based 
on the above study, we now realize that the violation of hermitian 
property is related to the violation of parity.  Though a non-
hermitian Hamiltoian is not allowed for a subluminal particle, 
it does work for tachyonic neutrinos. 

   As a tachyon, neutrinos have many peculiar features, which 
are very different from all other particles.  For instance,  
neutrinos only have weak interactions with other particles.
Neutrino has left-handed spin in any reference 
frame.  On the other hand, anti-neutrino always has 
right-handed spin.  This means that the speed of neutrinos
must be equal to or greater than the speed of light. Otherwise,
the spin direction of neutrino would be changed in some
reference frames.  Moreover, the energy of a tachyonic
neutrino (or anti-neutrino), $E_\nu$, could be negative in 
some reference frames.  We will discuss the subject of the 
negative energy in another paper. 
   
 The electron neutrino and the muon neutrino may have different
non-zero proper masses.  If taking the data from Ref.[16], then 
we obtain $m_s(\nu_e)=1.6 eV$ and $m_s(\nu_{\mu})=0.13 MeV$.
In this way, we can get a natural explanation why the numbers 
of e-lepton and $\mu$-lepton are conserved respectively.

 According to special relativity [23], if there is a superluminal 
particle, it might travel backward in time.  However, a re-
interpretation rule has been introduced since the Sixties [8-10].  
Another approach is to introduce a kinematic time under a non-
standard form of the Lorentz transformation [24-28].  Therefore, 
special relativity can be extended to space-like region, and 
tachyons are allowed without causality violation.

   Generally speaking, the above spin-${\frac{1}{2}}$ tachyonic 
quantum theory provides a theoretical framework to study the
hypothesis that neutrinos are tachyonic fermions.  More
measurements on the cosmic ray at the spectrum knee and more 
accurate tritium beta decay experiments are needed to further 
test the above theory.\\

\noindent
{\bf APPENDIX. EXPLICIT FORM OF SOLUTIONS}

  For a free particle with momentum $\vec p$ in the $z$ direction,
the plane wave can be represented by
 $$ \Psi(z,t)=\psi_{\sigma}exp[\frac{i}{\hbar}(pz-Et)]  $$
where $\psi_{\sigma}$ is a four-component bispinor.  This bispinor
satisfies the wave equation (7).  From Eq. (13), the explicit form
of two bispinors with the positive-energy states are listed as 
follows:
$$ \psi_1=\psi_{\uparrow (+)} = N \left(\matrix{1\cr
     0\cr a \cr 0 \cr}\right), \quad
    \psi_2= \psi_{\downarrow (+)}  = N \left(\matrix{0\cr
                       -a \cr 0 \cr 1 \cr}\right)    \eqno (A1) $$
and other two bispinors with the negative-energy states are:
$$ \psi_3=\psi_{\uparrow (-)} = N \left(\matrix{1\cr
     0 \cr -a \cr 0 \cr}\right), \quad
    \psi_4= \psi_{\downarrow (-)}  = N \left(\matrix{0\cr
                       a \cr 0 \cr 1 \cr}\right)    \eqno (A2) $$
where the component $a$ and the normalization factor $N$ are
$$  a=\frac{cp-m_s c^2}{|E|},  \quad
    N=\sqrt{\frac{p+m_s c}{2m_s c}}        \eqno (A3)  $$
For $ \psi_1=\psi_{\uparrow (+)}$, the conserved current becomes:
$$ \rho = \Psi_1{^\dag} \gamma_5 \Psi_1=\frac{|E|}{m_s c^2}, \quad
     j = \frac{p}{m_s}     \eqno (A4)  $$
and we can also obtain a scalar:
$$  {\bar\Psi_1} \Psi_1 = \Psi_1^{\dag} \beta \Psi_1 = 1 
   \eqno (A5)  $$

\vskip 0.8cm
  The author is grateful to G-j. Ni and Y. Takahashi 
for helpful discussions.  \\


\baselineskip=14pt
\noindent
\begin{thebibliography}{99}

\bibitem{1} R. Ehrlich, Phys. Lett. B {\bf 493}, (2000) 1.
Also see: LANL preprint hep-ph/000940.

\bibitem{2} R. Ehrlich, Phys. Rev. D {\bf 60}, (1999) 17302.

\bibitem{3} R. Ehrlich, Phys. Rev. D {\bf 60}, (1999) 73005.

\bibitem{4} Ch. Weinhermer et al., Phys. Lett. B {\bf 460} 
(1999) 219.

\bibitem{5} V.M. Lobashev et al., Phys. Lett. B {\bf 460} 
(1999) 227.

\bibitem{6} J. Bonn and Ch. Weinheimer, Acta Phys. Pol., {\bf 31}
(2000) 1209.

\bibitem{7} K. Assamagan et al., Phys. Rev. D {\bf 53} 
(1996) 6065.

\bibitem{8} O. M. P. Bilaniuk, V. K. Deshpande, and E. C. G. 
Sudarshan, Am.J.Phys. {\bf 30} (1962) 718.

\bibitem{9} E. Recami et al, Tachyons, Monopoles and Related 
  Topics, North-Holland, (1978), and references therein.

\bibitem{10} G. Feinberg, Phys. Rev. {\bf 159} (1967) 1089.

\bibitem{11} A. Chodos, A.I. Hauser, and V. A. Kostelecky, Phys.
Lett. B {\bf 150} (1985) 295.

\bibitem{12} T. Chang, "Does a free tachyon exist ?", 
Proceedings of the Sir A. Eddington Centenary Symposium, Vol. 3, 
Gravitational Radiation and Relativity", p.431 (1986). 

\bibitem{Chodos92} A. Chodos, V. A. Kostelecky, R. Potting, 
and E. Gates, Mod. Phys. Lett. A {\bf 7} (1992) 467.

\bibitem{Chodos94} A. Chodos, and V. A. Kostelecky, Phys.Lett. B 
{\bf 336} (1994) 295.

\bibitem{15} See e.g. E.C.G. Sudarshan: in Proceedings of the VIII 
Nobel Symposium, ed. by  N. Swartholm ( J. Wiley, New York, 1970), 
P.335;  J. Bandukwala and D. Shay,  Phys. Rev. D {\bf 9} (1974) 
889;  D. Shay, Lett. Nuovo Cim. {\bf 19} (1977) 333. 

\bibitem{16}  ''Review of Particle Physics'', Euro. Phys. 
Journ. C {\bf 15} (2000) 350.

\bibitem{17} P.A.M. Dirac, Proc. R. Soc. Ser, A {\bf 117}; 610, 
{\bf 118} (1928) 351.

\bibitem{18} T. D. Lee and C. N. Yang, Phys. Rev. {\bf 104} 
(1956) 254; Phys. Rev. {\bf 105} (1957) 1671.

\bibitem{19} C.S. Wu et al., Phys. Rev. {\bf 105} (1957) 1413.

\bibitem{20} G-j. Ni et al, Chin. Phys. Lett., {\bf 17} (2000) 393.

\bibitem{21} T. Chang and G-j. Ni, "An Explanation on Negative
Mass-Square of Neutrinos", LANL preprint hep-ph/0009291 (2000).

\bibitem{22} G-j. Ni and T. Chang, "Is Neutrino a superluminal 
particle ?", preprint Fudan0930 (2000). 

\bibitem{23} A. Einstein, H.A. Lorentz, H. Minkowski, 
and H. Weyl, The Principle of Relativity (collected papers), 
Dover, New York (1952).

\bibitem{24} R. Tangherlini, Nuov. Cim Suppl., {\bf 20} (1961) 1.

\bibitem{25} T. Chang, J. Phys. A {\bf 12}(1979) L203. 

\bibitem{26} J. Rembielinski, Phys. Lett., A {\bf 78} (1980) 33;
Int. J. Mod. Phys.,A {\bf 12} (1997) 1677.

\bibitem{27} T. Chang and D.G. Torr, Found. Phys. Lett., 
{\bf 1} (1988) 343.

\bibitem{28} P. Caban and J. Rembielinski, Phys. Rev., A {\bf 59} 
(1999) 4187.

\end{thebibliography}
\end{document}